\documentstyle[emulateapj,psfig]{article}
\begin{document}

\title{Asymmetric core combustion in neutron stars and a potential mechanism for
gamma ray bursts}
\author{G. Lugones,  C. R. Ghezzi,  E. M. de Gouveia Dal Pino and J. E. Horvath}
\affil{Instituto Astron\^omico e Geof\'{\i}sico, Universidade de S\~ao Paulo, \\
Rua do Mat\~ao 1226 - Cidade Universit\'aria, 05508-900 S\~ao
Paulo SP, Brazil \\ Email: glugones@astro.iag.usp.br}

\begin{abstract}
We study the transition of nuclear matter to strange quark matter (SQM) inside
neutron stars (NSs). It is shown that the influence of the magnetic field
expected to be present in NS interiors  has a
dramatic effect on the propagation of a laminar deflagration (widely studied so
far), generating a strong
acceleration of the flame in the polar direction.
This results in a strong asymmetry in the geometry of the just formed core of
hot SQM which  resembles a cylinder
orientated in the direction of the magnetic poles of the NS.
This geometrical asymmetry gives rise to a bipolar emission of the thermal
neutrino-antineutrino  pairs
produced in the process of SQM formation. The $\nu {\bar \nu}$  annihilate into
$e^+ e^-$ pairs
just above the polar caps of the NS giving  rise to a relativistic fireball,
thus providing
a suitable form of energy transport and conversion to $\gamma$-emission that may
be associated to short gamma ray bursts (GRBs).
\end{abstract}


\section{Introduction}

The transition to SQM is expected to occur in NSs if SQM has lower energy
per baryon than ordinary nuclear matter
\cite{bodmer71,terazawa79,chin79,witten84}.
This has strong consequences in the astrophysics of compact stars since,
if SQM is the true ground state of strongly interacting matter rather than
$^{56}$Fe,  compact objects
could be strange stars (SS) instead of NSs.
Some tentative strange star candidates are
the compact objects associated with the X-ray bursters GRO J1744-28
\cite{Cheng98},
SAX J1808.4-3658 \cite{Li99a}, and  the  X-ray pulsar Her X-1  \cite{Dey98}.
In addition, the observed high and
low frequency  quasi periodic oscillations in the atoll source 4U 1728-34
have been shown to be more consistent with a strange star nature \cite{Li99b}.
A number of different mechanisms has been proposed for  NM  to SQM
conversion \cite{aarhus} inside the star.  All them are based on the formation
of a ``seed'' of SQM
inside the NS. A possible mechanism is the so-called strangelet contamination,
where a seed of SQM from the ISM enters a NS and converts it to a SS.
In another possible scenario,  a seed of SQM forms in the core of a NS as a
result of the increase of the central density above the critical density for
deconfinement
phase transition. A way to do this is through mass accretion onto the NS in a
binary stellar system.
In a third mechanism, a seed of SQM may naturally form inside a newly born
neutron
star from a core-collapse supernova explosion after a deleptonization timescale
\cite{mnras1,mnras2}.
No matter which of these mechanisms are actually triggering the NS to SS
transition, once
the first seed of SQM is produced inside the NS, it will propagate as a
combustion swallowing
neutrons, protons and hyperons.
The transition to SQM inside the burning front has actually two stages.
Deconfinement
driven by {\it strong} interactions first liberates quarks
confined inside hadrons. The just deconfined quark phase has a certain finite
strangeness due to the presence
of strange hadrons in NS matter.
However, the composition of the just deconfined phase (with u,d and s quarks) is
not in beta
equilibrium and consequently chemical equilibrium is reached by {\it weak}
interactions.
It is in this phase that a large amount of energy is released and then $\nu$'s
are copiously
produced.

%---------------------

The type of diffusion-driven combustions studied so far by several authors
\cite{Slow1,Slow2} actually correspond to laminar deflagrations (slow
combustions).
Whether the conversion process remains forever as a deflagration, either laminar
or turbulent;
or jumps to the detonation regime, thus driving an explosive transient
\cite{Comb1,Comb3,Comb2} has been debated
in the literature. The situation is closely analogue to the much more studied
thermonuclear
combustions leading to type Ia supernovae.
In addition to the lack of detailed studies on the mode(s) of
combustion (laminar/turbulent deflagration or detonation), there is (to the
best of our knowledge) no calculation of the influence of ubiquitous
magnetic fields expected to be promptly generated  or already present
in the NS as a fossil. We shall consider hereafter
an initially laminar deflagration in presence of a
$B$-field. As stated, previous works  \cite{Slow1,Slow2} have calculated the
velocity of the laminar deflagration by considering the diffusion of
$s$-quarks as the main agent for the progress of the conversion.
Given that the combustion is idealized to happen near the $T = 0$
limit (which is small when  compared with the chemical potential of quarks),
these
diffusion-limited "cold" models are reasonable for this purpose.
The general result, which also holds when temperature corrections
and full non-linearity are considered, is that the laminar velocity
of the front is relatively slow ($v_{lam} \leq 10^{4} cm/s$) although this speed
may be uncertain by several orders of magnitude (Olinto 1991).
This velocity is a direct consequence of  both the timescale for weak
decays that create $s$-quarks ($10^{-8} s$)
and the physics of the diffusion process.

\section{Instabilities and asymmetries}

A nuclear flame that starts as a laminar deflagration propagating outward
rapidly enters the wrinkled flamelet regime due to the action of
several hydrodynamic instabilities,
such as the Landau-Darrieus (LD) and Rayleigh-Taylor (RT).
For the largest scales (those $\gg$ than the thickness of the flame),
the RT instability  dominates over LD (see Ghezzi et al. 2001 for details).
It is possible to identify distinct regimes in the deflagration stage,
and we will apply the fractal model of combustion \cite{f1,f3,f4,Ghezzi} for
all those regimes.
The wrinkled surface $\bar A$ behaves like a fractal with
$\bar A \propto \bar{R}^{D}$,  where $D$ is the fractal dimension of the
surface, with
$2\leq D<3$, and $\bar{R}$ is the mean radius of the wrinkled surface
\cite{Filyand}.
Numerical simulations \cite{Filyand} and laboratory experiments involving
different
gas mixtures \cite{Gostintsev} show that the fractal growth
actually increases the velocity of the combustion front because of the change in
the transport mechanism from a laminar to a fully turbulent burning.
The burning of nuclear matter to SQM has been studied so far in the absence of
any
$B$-field. However, there is an ample consensus that
strong magnetic fields should be
present at the center, as indicated by the observations of
surface pulsar fields  ( $B \, \geq \, 10^{11-12} \, G$ ). Even though there is
no detailed
information on the fields of non-pulsating NSs,
it seems reasonable to assume a field at least as strong as the one at
the surface of radio pulsars {\it inside} the object.

%------------------------------------------------------------

The fractal model presented in Ghezzi et al. 2001 for the
thermonuclear burning of a white dwarf is applicable to the
present case since all the basic ingredients and the key points of
the physics are similar. For the sake of simplicity, we shall assume hereafter a
dipolar geometry and thus the flame propagating in two particularly
representative directions, one parallel to the B-field lines (in the polar
direction) and the other one perpendicular to them (in the equatorial
direction).
Within the fractal description,
the effective velocity of the flame  is
given by $v_{f} = v_{lam} ( L /l)^{D-2}$ where  L and
$l$ are the maximum an minimum length scales of
perturbations unstable to RT .
The characteristic velocity of the RT growing modes must be
$\ge v_{lam}$, i. e.  $l n_{RT}(l) =
v_{lam}$, where  $n_{RT}$ is the inverse of the characteristic RT
time. In the absence of a B-field $n_{RT}= (1/2\pi) \sqrt{gk
\Delta \rho /2 \rho}$ and so $l = 4 \pi \rho v^2_{lam}/g
\Delta \rho$. However, in the equatorial direction, the presence
of B is essential in modifying  the dispersion relation for the RT
instability which reads $n_{RTB} =1/ 2\pi \sqrt{ gk (\Delta \rho
/2 \rho - k B^2/4 \pi g \rho ) }$
where  $k= 2 \pi /\lambda$, $\lambda$ is the wavelength of the perturbation,
and $\Delta\rho = \rho_u - \rho_d$ is the density difference  between
the upstream and downstream parts of the flame front \cite{Ghezzi}.
Taking into account the velocity of the flame, the minimum cut-off
length is given by the condition $l_{e} n_{RTB}(l_{e}) =
v_{lam}$, from which we derive the minimum scale in the equatorial
direction $l_{e} = 8 \pi( B^2/8 \pi +   \rho v^2_{lam}/2  ) / g
\Delta \rho$. So, from the above definition of the fractal
velocity we have $v_{e} = v_{lam} ( L /l_{e})^{D-2}$ and
$v_{p} = v_{lam} ( L /l_{p})^{D-2}$. Note that L is not
modified by B and has the same value in both directions. Then,
the ratio between the equatorial and polar velocities is
\cite{Ghezzi}:
\begin{equation}
\xi  = v_p/v_e = [ 1 + B^2 / ( 4 \pi {\rho}  \, v_{lam}^{2} )
]^{D-2} .
\end{equation}

\noindent That is, although the magnetic pressure $B^2 /(8 \pi)$ is not
relevant, the  B-field quenches the growth of RT
instabilities in the equatorial direction acting as a surface
tension, while it is innocuous in the polar one where in average we
have $\vec{v}_{p} \times \vec{B} = 0$ . More specifically  B modifies the
minimum RT instability scale and since the turbulent flame
velocity is related to RT-growth this results in a different
velocity of propagation along each direction.
Since the density of the products of combustion (SQM) is comparable to the
fuel (nuclear matter) $\rho \sim 10^{15} g/cm^{3}$;  and using the values for
$v_{lam}$
found in previous works \cite{Slow1,Slow2} ($\sim 10^{4} cm/s$), we find that
$B^{2}/(4 \pi) > {\rho} v_{lam}^{2}$ for
relatively low values of $B (i.e. \sim 10^{12-13} G$), so that
for $D \approx 2.5$ (which is in good agreement with numerical studies,
Blinnikov et al. 1995),
$\xi$ scales linearly with the field to a good approximation. So,
\begin{equation}
\xi \simeq  10 \times  (B/10^{13} G) (10^{15}g cm^{-3}
/{\rho})^{1/2} (10^{4} cm s^{-1} /v_{lam})
\end{equation}

\noindent and large asymmetries can be produced even for moderate
values of $B$ \footnote{We should note that, although the present work is based
solely on a linear analysis, 3D numerical simulations of the development of the
magnetic RT instability (Jun et al. 1995) reinforces the results of asymmetry
here reported and also reveals a tendency for its amplification in the non-
linear regime.}. Next, we evaluate the time that the polar front
needs to reach the surface $(\tau_{p} =R/v_p = R/(\xi v_e) )$:
\begin{equation}
\tau_{p} \simeq  10 s  \times  (R /10 km) ( 10^{13} G / B ) ( {
\rho} / 10^{15} g cm^{-3})^{1/2}. \label{tau}
\end{equation}

\noindent Note that eq. (3) gives actually an upper bound to $\tau_p$ because we
have approximated $v_e$ by $v_{lam}$. An asymmetry in the $\nu$-emission will be
possible only if
$\tau_{p}<\tau_{d}$,
that is, if $\tau_p$  is smaller than the typical diffusion timescale $\tau_{d}$
through nuclear matter
in the equatorial direction ($\tau_{d}$ is at least $30 s$, Pons et al. 1999).
Only in this case we can assume
that the lateral sides of the
cylinder are opaque and almost all the $\nu \bar{\nu}$ are emitted through the
polar caps.
Otherwise, we will still have a strong asymmetry
in the shape of the SQM region but this will not yield an asymmetry in the
$\nu$-emission structure. The minimum $B$ for which
$\tau_{p}<\tau_{d}$ is satisfied is obtained from Eq. (\ref{tau})
\begin{equation}
B_{min} =  3 \times 10^{12} G  ( { \rho} / 10^{15} g cm^{-3}
)^{1/2} ( R / 10 km ).
\end{equation}

\noindent In other words, for $B < B_{min}$ we recover the isotropy of the
neutrino emission (see below).

When the polar front reaches the stellar surface, the equatorial front will have
travelled only
$\delta = R/ \xi = 1  km   \times (R/10 km) (10^{13} G / B) ({\rho} / 10^{15} g
cm^{-3})^{1/2} (v_{lam} /  10^{4} cm s^{-1})$,
and therefore the SQM region will begin to emit its neutrino content
mainly through a cylinder aligned with the poles
with a radius $\delta \simeq 1$ km and a height $2 R \simeq 20$ km (see Fig 1).
Notice that if $B \gg 10^{13}G$, $\delta$ is very small and the total energy
released by the SQM conversion is negligible in this asymmetric stage.
An estimate for the ratio of the neutrino flux in the equatorial and the polar
directions is found  by comparing the flux through the opaque NS matter in the
equatorial direction with the free streaming flux of neutrinos emitted from the
polar caps, namely
\begin{equation}
F^e / F^p =  [(-c \lambda_n / 3)  \; \partial \epsilon_{\nu} /
\partial r  ] \; / \; [ c  \epsilon_{\nu}]
\approx   \lambda_n / (3 \Delta r)
\end{equation}

\noindent where  $\Delta r$  is the difference between the radius of the star
and the equatorial radius of the SQM core.
$\lambda_{n} = 3 \times 10^{3}  cm ( \rho_{nuc} / \rho) (10 MeV / E_{\nu})^2 $
is the mean free path of neutrinos diffusing  from the surface of the SQM core
in the equatorial
direction \cite{Shapiro}.  This $\lambda_{n}$ is mainly due to neutral-current
elastic scattering off neutrons.
For typical values of $\lambda_{n}$  and  a highly asymmetric SQM region we have
$F^e/F^p \sim 10^{-2} - 10^{-4}$. Thus, once the polar front reaches the
surface,
almost all $\nu$'s are released through the polar caps.
This is due to the fact that the flux in the equatorial direction is quenched
due to the high $\lambda_n$
while $\nu {\bar \nu}$ 's  in the polar surface escape freely.

The neutrino transport inside the SQM cylinder can be described by:
${\partial e_{tot}}/{ \partial t} = {\partial}/{ \partial x} [(\lambda_{SQM}
c/3) \;  {\partial e_{\nu}}/{ \partial x}]$,
with the boundary conditions ${\partial e_{\nu}(0,t)}/{ \partial x} = H
e_{\nu}(0,t)$,
${\partial e_{\nu}(l,t)}/{ \partial x} = H e_{\nu}(l,t)$
and the initial condition $e_{\nu}(x,0) = e_{\nu 0}$, being $e_{\nu}(x,t)$ the
neutrino energy density,
$e_{tot}(x,t)$ the total energy density of SQM, $l=2R$ the height of the
cylinder,
and $H= 3 / \lambda_{SQM}$. For the sake of simplicity, we assume that neutrino
transport is driven
by electron-neutrino pairs, so that $\lambda_{SQM} = 1.3 \times 10^{3} cm /
T_{11}^2$ \cite{Haensel}.
Although muon and tau neutrinos are in fact important, this simplification  does
not essentially modify
the picture and allows a simple solution  for  the  problem.
The total  internal energy released in the conversion of a whole NS into SQM is
$E_I \simeq {\rm few}  \times 10^{53}$
ergs \cite{Bombaci}. So, a rough estimate of the initial  temperature of the
just formed SQM will be $T_{11} \simeq 3$ (in units of $10^{11} K$).
From the integration of the equations above we obtain the temperature in the
polar cap surface as a function
of time $ T_{p11}(t)$, and the corresponding neutrino luminosity
\begin{equation}
L_{\nu}(t)  = 1.2 \times 10^{53}  erg/s  \;\; [T_{p11}(t) / 3]^4 (
\delta / 1 km )^2.
\end{equation}

\noindent The emitted $\nu {\bar \nu}$'s   annihilate into  $e^+ e^-$ pairs with
rather low efficiencies.  Following the procedure of Haensel et al. 1991 we have
calculated
the luminosity in $e^+ e^-$  for this cylindrical configuration
(see Haensel et al. 1991 for the result in spherical symmetry)
\begin{equation}
L_{pair}(t)  = 2.1 \times 10^{52} erg/s \;\;  [T_{p11}(t) / 3]^9 \; ( \delta /
1 km )^2  ( R /10 km ). \label{7}
\end{equation}

\noindent These luminosities are shown in Fig. 2. We find that
$90 \%$  of the $e^+e^-$ pairs are injected inside  small
cylinders located just above the polar caps (with radius $\delta$
and height $0.4 R$) in a timescale of $\tau_i \simeq 0.2 s$ almost
independently on the initial temperature.

\section{Application to GRBs}
GRBs appear to fall  into at least two distinct categories, namely the short
duration bursts
($\approx$ 0.2 s) and the long duration ones ($\approx$ 20 s)
(for a review see \cite{Piran}).
Previous works have explored  the idea that the conversion of nuclear matter
into SQM
in NSs may be an energy source for GRBs
\cite{G1,G2,Haensel,G3,Bombaci,G4,G5}.
These models addressed
spherically symmetric conversions of the whole NS giving isotropic $\gamma$-
emission.
We show here that if a conversion to SQM actually begins near the center of a
NS, the presence of a {\it moderate} magnetic field $B$ ($\sim 10^{13} G$)
will  originate a prompt asymmetric
$\gamma$-emission which will be observed as a short, beamed GRB.
Given that all pairs contributing to Eq.(\ref{7}) will annihilate into
$\gamma$'s, we are lead to the conclusion that an asymmetric
short GRB can be generated by this process. The energy injected in $\gamma$-rays
is $E_{\gamma} \ge 10^{51}$ erg,
comparable to the ones derived for long GRBs.
The asymmetry crucial for this conclusion is controlled by the quotient of the
field $B$ to the laminar velocity
$v_{lam}$ (Eq.1). Had the latter been larger by a few orders of magnitude no
significant asymmetry would be produced by realistic magnetic fields.
It should be remarked that we have {\it not} shown that the
turbulent deflagration will be the actual propagation mode, but rather
assumed its realization all the way down to the stellar surface.
Actually, a possible transition from a deflagration to  a detonation
mode (DDT) (in close analogy with the DDT in white dwarf combustions),
may be achieved  \cite{Comb1,Comb3,Comb2}
depending on the behavior of the microphysics.
However,  if a substantial asymmetry has already been produced when the DDT
is achieved, the polar flow will detonate before the equatorial one.

Basic requirements for associating asymmetric SQM  burnings to (short)
GRBs are the absence of a large amount of matter above the polar regions
and a rate per galaxy high enough ($\sim 10^{-3} yr^{-1}$) so as to compensate
the collimation of the emission.
It is unclear whether both can be achieved, but in any case it is clear that
further
studies of the hydrodynamics and statistics of the events are desirable.
However, and quite independently of the specific astrophysical setting, the
result of NM to SQM conversion in presence
of  $B \sim 10^{13} G$ will produce an asymmetry in the  neutrino emission and
an enhanced annihilation to
$\gamma$'s.  However, the observational outcome and the rate of the events  will
depend on the type of system in which the NS is being burned.
For example, in transitions produced by accretion induced collapse of a white
dwarf,
the energy is deposited in a region that contains a low baryon loading.
So, this collimated emission has an advantage over the isotropic one, where
$B=0$,
in that the total baryon loading ``seen'' by the beamed burst
($ \sim \delta^2 /(4 R^2) =   \pi {\rho}  v_{lam}^2  / B^2 \simeq 10^{-2}-10^{-
3}$ of the total)
is low enough to allow for a relativistic expansion of the fireball with very
high Lorentz factors and is sufficient to explain apparent burst luminosities
$L_{\gamma}$   up to more than $  \sim 10^{52} $
ergs/s for burst durations of $t_{\gamma} \approx 0.2 s$.
NSs in Low Mass X-ray Binaries (LMXBs) are also likely
candidates for the conversion to SQM because mass accretion may raise the
central density of the NS above
the critical density for transition. These NSs have "weak"
magnetic fields in their surface (B = $10^{8 - 9}$ G) which
are generally understood  as the result of mass
accretion. Since this affects in principle only the field in the surface, the
field in the {\it interior} of the NS would be in the range of values that leads
to an asymmetric short GRB. Note also that if the emission is beamed it would be
difficult to explain the observed rate of GRBs by conversions in binary systems
alone.
For transitions in newly born proto-NSs \footnote{We should note that according
to recent calculations of  proto-NS evolution,  convective instabilities occur
within the first second of NS formation (Janka et al. 2001). Since the
transition to SQM is not expected to occur immediately after the proto-NS
formation, but after a sensible fraction of the deleptonization timescale
(Lugones and Benvenuto 1998, Benvenuto and Lugones 1999), it is very likely
that convection will have stopped well before the start of the burning,
therefore not affecting the asymmetric scenario described in this work.}, the
presence of
the supernova envelope
makes necessary  to explain  how the $\gamma$-emission produced
by the central proto-NS traverses the young expanding ejecta
(one clue may have been provided by GRB980425/SN1998bw association). In any
case, we may expect emission of quasi-thermal X-ray flashes with rates per
galaxy
$\eta 10^{-2} yr^{-1}$,  being $\eta$ the unknown
fraction of proto-NS with fields  $B \sim 10^{13} G$ from the events.
Other consequences for the ejected envelopes may include
asymmetries of the envelope itself due to the non-isotropic injection of energy.
We plan to discuss these issues thoroughly in a future work.

The FAPESP-S\~ao Paulo and CNPq Agencies are acknowledged
for partial financial support to all authors.

\newpage

\begin{figure}
%\plotone{f1.eps}
\caption{A {\it moderate} field ($B \sim10^{13}
G$) in the interior of a NS being burned to SQM is able to
generate an asymmetric velocity field in the phase change front
which, depending on the exact value of $B$, may reach values $\gg
1$. So, a prolate  SQM fluid is the actual geometry achieved by
the conversion in presence of such non-zero $B$. This  affects the
leaking of thermal neutrinos produced in the transition,
funnelling them into opposite polar cones.}
\end{figure}

\begin{figure}
%\plotone{f2.eps}
\caption{Neutrino and pair luminosities for $B=
10^{13} G$. The timescale on which 90 \% of the pairs are injected
is $0.2 s$.}
\end{figure}

\begin{figure}
\plotone{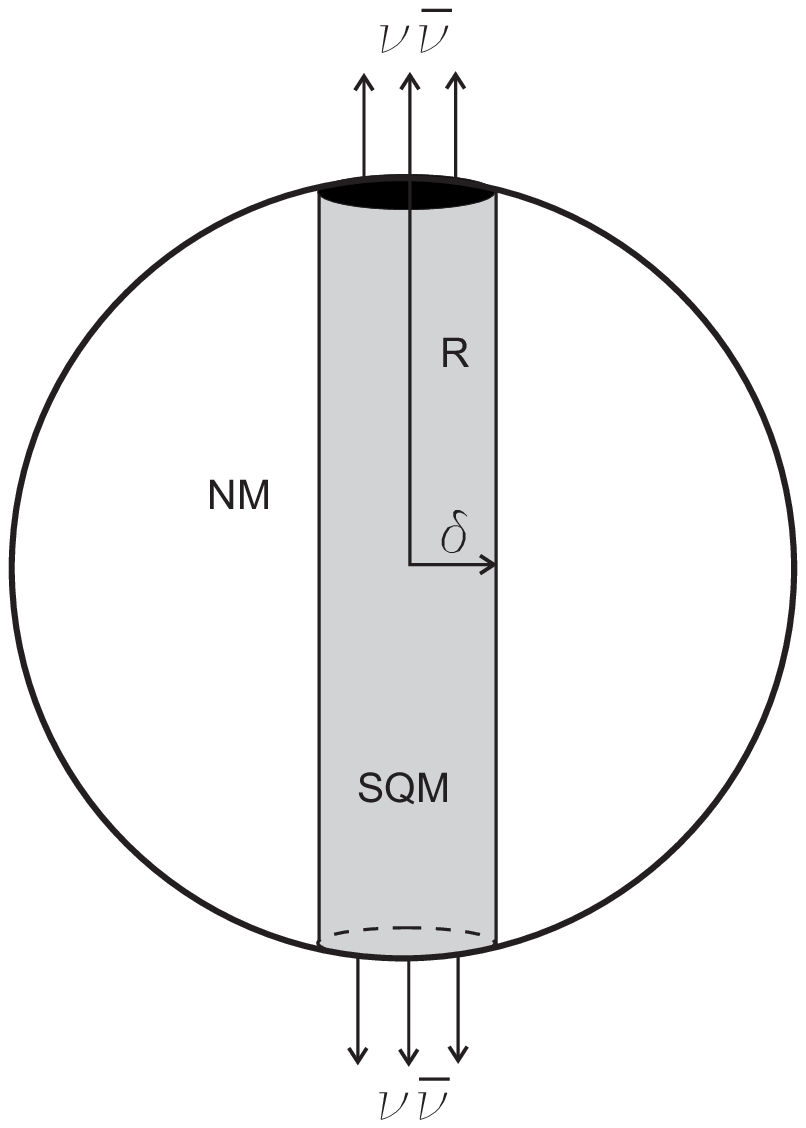}
\end{figure}

\begin{figure}
\plotone{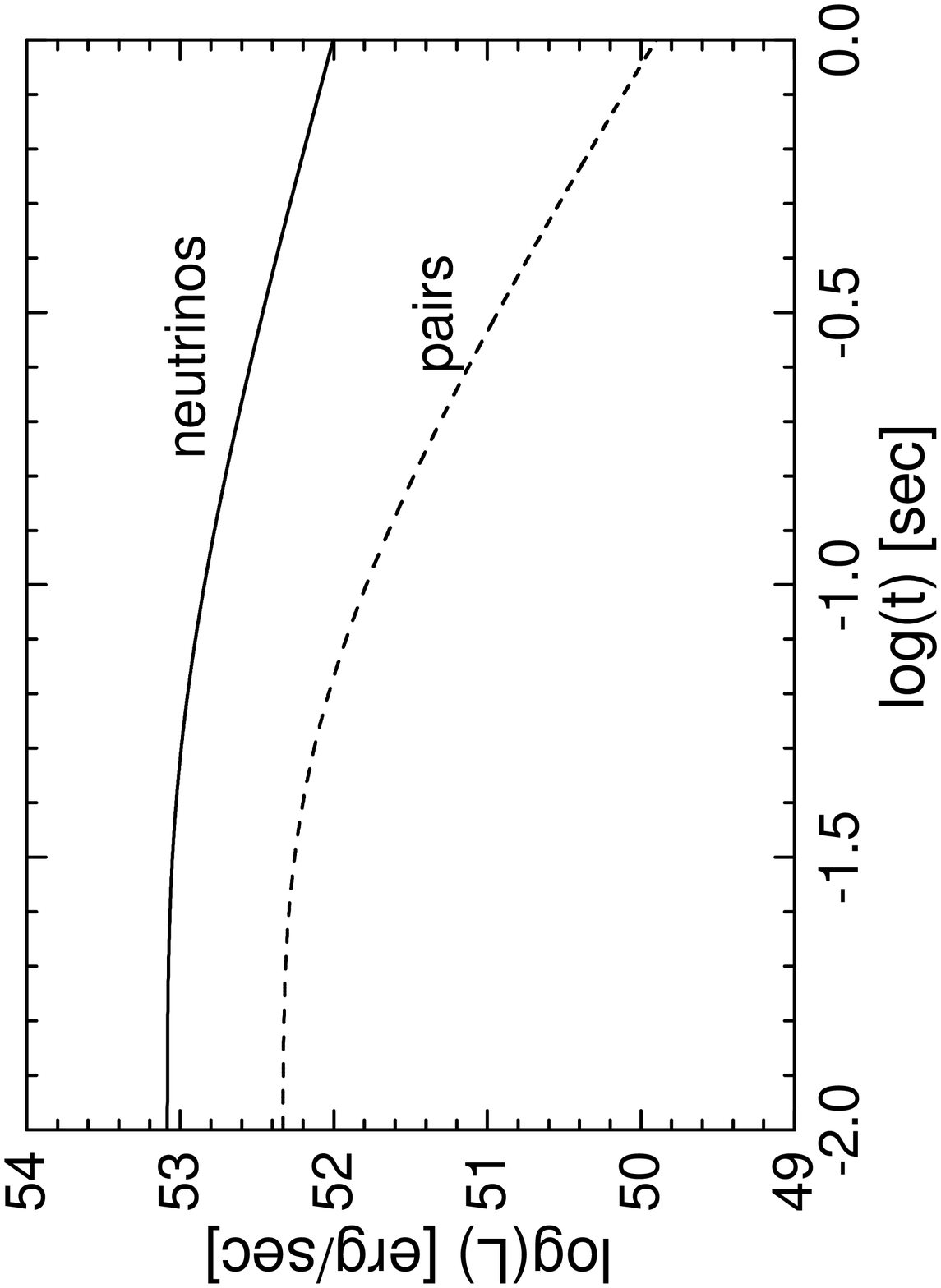}
\end{figure}

\end{document}